# ON THE SCALABILITY OF MULTIDIMENSIONAL DATABASES

István SZÉPKÚTI

ING Nationale-Nederlanden Hungary Insurance Co. Ltd.
Andrássy út 9, H-1061 Budapest, Hungary
e-mail: szepkuti@inf.u-szeged.hu

**Abstract**

It is commonly accepted in the practice of on-line analytical processing databases that the multidimensional database organization is less scalable than the relational one. It is easy to see that the size of the multidimensional organization may increase very quickly. For example, if we introduce one additional dimension, then the total number of possible cells will be at least doubled.

However, this reasoning does not take into account that the multidimensional organization can be compressed. There are compression techniques, which can remove all or at least a part of the empty cells from the multidimensional organization, while maintaining a good retrieval performance. Relational databases often use B-tree indices to speed up the access to given rows of tables. It can be proven, under some reasonable assumptions, that the total size of the table and the B-tree index is bigger than a compressed multidimensional representation. This implies that the compressed array results in a smaller database and faster access at the same time.

This paper compares several compression techniques and shows when we should and should not apply compressed arrays instead of relational tables.

*Keywords:* scalability, multidimensional database, On-line Analytical Processing, OLAP.

## 1 Introduction

### 1.1 Motivation

The total number of cells in multidimensional matrices or arrays increase quickly. Consider an n-dimensional array and assume that we want to add a new dimension to it. The new dimension will contain at least two elements, otherwise the cells of the new array will not depend on the new dimension. Let $c_{n+1} \geq 2$ denote the number of dimension values in the new dimension ($c_{n+1} := |D_{n+1}|$). Then, the total number of cells in the new array will be $c_{n+1}$ times more than in the old array, it will equal the following expression

$$|D_1 \times D_2 \times ... \times D_n \times D_{n+1}| \qquad (1)$$

where $D_i$ denotes the set of dimension values of the $i^{th}$ dimension ($i = 1, 2, ..., n, n+1$). This means an exponential increase in the total number of cells with the increase of the number of dimensions, given that $c_i := |D_i| \geq 2$ for all $i$. That is why one may conjecture that the multidimensional databases are not scalable.

Fortunately, we do not have to store all the possible cells. There are several compression techniques, which are able to remove at least a part of the empty cells. The intention of this paper is to prove the existence of such compression techniques, which result in a smaller database than the relational (that is tables-based) database organization while maintaining a faster retrieval performance.

## 1.2 Scalability

There are a lot of different possible definitions of scalability. For instance Pendse mentions eleven types of scalability in [6]. These *scalability types* and their descriptions are the following:

> *Data volumes.* Ability to handle very large volumes of input data (excluding pre-calculated results, indexes, metadata and other overheads), with acceptable load/calculate/query performance. [Acceptable query response in very large databases may be minutes, whereas it could be sub-second in very small databases.]
>
> *Dimension size.* Large numbers of members in a single hierarchical dimension. Includes the capability to administer and update large dimensions and databases using them.
>
> *Dimensionality.* Large numbers of base and/or virtual independent dimensions.
>
> *Dimensional model.* Multiple, linked cubes of varying dimensionality.
>
> *Numbers of users.* Performance, capacity, security and administrative features to support large numbers of concurrent users plus administrative capabilities for larger numbers of registered users.
>
> *Calculations.* Sophisticated multidimensional calculations.
>
> *Platforms.* Ability to run on multiple server and client platforms.
>
> *Functionality.* Full range of capabilities to implement many different types of applications.

*Deployability.* Ability to deploy acceptable business solutions without excessive implementation services.

*Affordability.* Cost effective hardware, software, implementation.

*Front-end options.* Option to use multiple vendor and third-party front-end tools to suit different user types.

Throughout this paper, we are going to take the database size as the primary criterion for scalability.

## 1.3 Results

The results of this paper can be summarized as follows:

- It introduces two improvements of the single count header compression scheme: the logical position compression and the base-offset compression.

- It proves theoretically that, in the worst case, the multidimensional array representation with single count header compression occupies less space than the table-based database organization with B-tree.

- Using two benchmark databases (TPC-D and APB-1), it proves empirically that size of the multidimensional representation with base-offset compression is smaller than the size of the table representation with B-tree, even if the size of the B-tree is minimal.

- In case of the TPC-D benchmark database, the size of the multidimensional representation with base-offset compression is smaller than the size of the compressed table representation with B-tree.

- In addition to the size advantages, the paper also proves with experiments that the retrieval operation is 1.4 - 7.8 times faster in the multidimensional representation than in the table representation.

## 1.4 Related Work

The list of scalability types comes from the paper of Pendse [6].

There are several compression techniques in the literature of on-line analytical processing databases.

The conjoint dimension appeared in Express. It was the first multidimensional analytical tool and dates back to 1970 [5]. Now, it is a product of Oracle.

The paper of Zhao et al. [9] introduced the chunk-offset compression. The authors of [9] performed extensive experimentation. They compared, among others, the performance of the cube operator on relational that is table-based (ROLAP) as well as multidimensional (MOLAP) database organization. The cube is an aggregation operator, which generalizes the group-by clause of an SQL statement. They found that their Multi-Way Array method performs much better than the previously published ROLAP algorithms. Moreover, the performance benefits of the Multi-Way Array method are so substantial that in their tests it was faster to load an array from a table, cube the array, then dump the cubed array into tables, than it was to cube the table directly. In [9], the cube operator was examined, whereas in this paper retrieval is tested. It is worth to note the similarities, as well. In [9], just like in this paper, the compressed multidimensional array occupied less space than the table representation; and at the same time, the compressed multidimensional array results in faster operation than the table-based physical representation. The retrieval operation is interesting because for example the more complicated aggregation operation, uses it in bulk. Therefore, faster retrieval may imply faster aggregation. The experimentation results of [9] supports this statement.

The single count header compression scheme is described in [1]. We used a variation of this scheme: instead of storing the cumulated run lengths of empty and nonempty cells, we store the $(L_j, V_j)$ pairs of logical positions and number of empty cells up to the logical positions ($L_j$). This variation was described in [7] for the first time.

The Theorem in section **Comparison of Physical Database Representations** may be related to Assertion 2 in [7]. This latter assertion says that the multidimensional representation occupies less space than the table representation, if the data ratio ($\delta$) of the table is smaller than the density ($\rho$) of the multidimensional array. The difference between the Theorem and Assertion 2 is threefold:

- Assertion 2 does not take any compression into account, whereas the Theorem assumes single count header compression.

- That is why Assertion 2 disregards the size of the B-tree in the table representation and the size of the header in the multidimensional array representation.

- The Theorem is proved in the worst case scenario, without any other conditions, whereas Assertion 2 gives a sufficient condition.

## 1.5 Paper Organization

The rest of the paper is organized as follows. Section 2 describes compression techniques. Section 3 improves on single count header compression. Section 4 proves that a variation of single count header compression occupies less space than the table-based database organization. Section 5 gives the results of the experimentation. After the Conclusion, the paper ends with the Acknowledgments, the Appendix and the References.

## 2 Compression Techniques

There are a lot of compression techniques in the literature. In this section we are going to show only three of them.

*Conjoint dimension.* Let us suppose that the finite relation $R \subseteq D_1 \times ... \times D_n$ has a special property: given elements of $D_1 \times ... \times D_h$ ($1 \leq h \leq k \leq n$ and the unique primary key of R is constituted by $D_1, ..., D_k$) cannot be found in the corresponding projection of R. Thus, in order to eliminate empty cells form the multidimensional array representation, we can define an equivalent $R'$ relation:

$$R' = \{((d_1, ..., d_h), d_{h+1}, ..., d_n) \mid ((d_1, ..., d_h), d_{h+1}, ..., d_n) \in Conjoint \times D_{h+1} \times ... \times D_n \text{ such that } (d_1, ..., d_h, d_{h+1}, ..., d_n) \in R\}$$

where

$$Conjoint = \pi_{D_1, ..., D_h}(R)$$

Here, $\pi$ denotes the projection operation of relations.

**Definition.** Let us introduce the density of the multidimensional array $0 < \rho \leq 1$:

$$\rho = \frac{number\ of\ nonempty\ cells}{total\ number\ of\ cells} = \frac{|R|}{|D_1 \times ... \times D_k|} \qquad (2)$$

$\square$

Let $\rho'$ denote the density of the previously defined relation $R'$. Obviously, the density will decrease according to the following formula:

$$\rho' = \frac{|D_1 \times ... \times D_h|}{|Conjoint|}\rho \qquad (3)$$

We have to be careful with conjoint dimensions. Consider, for example, the case when h = k, that is all elements of the unique primary key are put into Conjoint. One can see that we could eliminate all empty cells this way and the multidimensional representation became identical with the table-based one. (The multidimensional representation of a relation is a multidimensional array or matrix, whereas the table-based representation is nothing else than a table in a relational database.) Thus, we have to exclude this extreme case of Conjoint, because it probably degrades the performance.

*Chunk-offset compression.* First, the n-dimensional array is divided into small size n-dimensional chunks. Then, the dense chunks (where the density $\rho > 40\%$) are stored without any modification. Sparse chunks are condensed using "chunk-offset compression." The essence of this method is that only the existing data are stored using (offsetInChunk, data) pairs. Within the chunk, the offset is calculated as follows:

$$i = ((...((i_k - 1)c_{k-1} + i_{k-1} - 1)...)c_2 + i_2 - 1)c_1 + i_1 \qquad (4)$$

where the calculated $i$ is called one-dimensional index, within the multidimensional index $(i_1, ..., i_k)$, $i_j$ denotes the index of dimension $D_j$ and $c_j = |D_j|$ is the number of dimension values in dimension $D_j$ ($1 \leq j \leq k$).

In this compression method, not all the sparse cells are removed from the array. In the pessimistic scenario, when all chunks are just slightly denser than 40%, almost 2.5 times more space is needed to store the cell values, because all empty cells are also stored in this case. This may result in up to 2.5 times more disk input/output operation than absolutely necessary, when the chunks are read or written.

*Single count header compression.* By transforming the multidimensional array into a one-dimensional array, we get a sequence of empty and nonempty cells:

$$(E^*F^*)^* \qquad (5)$$

In the above regular expression, E is an empty cell and F is a nonempty one. The single count header compression (SCHC) stores only the nonempty cells and the cumulated run lengths of empty cells and nonempty cells. In [7], we used a variation of the SCHC. The difference between the two methods is that the original method accumulates the number of empty cells and the number of nonempty cells separately. These accumulated values are stored in a single alternating sequence. The sum of two consecutive values corresponds to a logical position. (The logical position is the position of the cell in the multidimensional array before compression. The physical position is the position of the cell in the compressed array.) Thus, we have to look for a given logical position between these sums. In [7], instead of storing a sequence of values, we chose to store pairs of logical positions and number of empty cells up to this logical position: $(L_j, V_j)$. Searching can be done directly on the $L_j$ values; we do not have to sum two consecutive values of a sequence. This results in a simpler searching algorithm, when we want to do logical-to-physical position transformation. On the other hand, if one has to determine the physical position from an $(L_j, V_j)$ pair, then he or she has to take the difference $L_j - V_j$. In case of the original method, this physical position is explicitly stored; it is nothing else than the accumulated number of nonempty cells. Therefore, the implementation of the physical-to-logical position conversion may be simpler with the original SCHC. In the rest of the paper, when we mention SCHC, we refer to the variation of this compression scheme defined in [7].

**Definition.** The array storing the $(L_j, V_j)$ pairs of logical positions and number of empty cells will be called the SCHC header. □

## 3 Improvements

This section gives the description of two compression techniques, which improve on SCHC, if the SCHC header is maximal.

*Logical position compression.* By mapping the $(i_1, ..., i_k)$ k-dimensional index into a one-dimensional index, we can create a one-dimensional array (sequence) from the k-dimensional array:

$$(E^*F^*)^* \tag{6}$$

The meaning of E and F is just the same as in the previous section.

The size of the SCHC header depends on the number of $E^*F^*$ runs. In the worst case, there are $N = |R|$ runs. Then the size of the SCHC header is $2N\iota$. (We assume that $L_j$ and $V_j$ are of the same data type and each of them occupy $\iota$ bytes of memory.) But in this case, it is better to build another type of header. Instead of storing the $(L_j, V_j)$ pairs, it is more beneficial to store only the $L_j$ sequence of all cells (that is not only the $L_j$ sequence of runs).

The physical-to-logical position conversion is done through a simple binary search. The physical position $P(L)$ of logical position $L$ is defined as follows:

The physical position $P(L)$

- equals $j$, if there exists $L_j$ such that $L = L_j$;
- is undefined, otherwise.

$P(L)$ is undefined if and only if the cell at logical position $L$ is empty. The physical-to-logical position conversion if just a simple lookup of an array element:

$L(P) = L_P$

where $L(P)$ denotes the logical position of physical position $P$.

**Definition.** The compression method, which uses the sequence of logical positions only, will be called logical position compression (LPC). The $L_j$ sequence used in logical position compression will be called LPC header. □

The number of $E^*F^*$ runs is between 1 and $N = |R|$. Let $\nu$ denote the number of runs. Because the size of $L_j$ and $V_j$ is the same, the header is smaller with logical position compression, if $\frac{N}{2} < \nu$. Otherwise, if $\frac{N}{2} \geq \nu$, the logical position compression does not result in smaller header than the single count header compression.

The header with logical position compression is half of the SCHC header in the worst case, that is when $\nu = N$. Almost this is the case in the TPC-D benchmark database: $\nu = 6,000,568$, whereas $N = 6,000,965$. (For the description of the relation, which we refer to here, see the section entitled **Experiments**. The TPC-D benchmark database itself is specified in [8].)

If the header is halved approximately, then a bigger portion of the header fits into the memory. In a virtual memory environment, this means less paging and thus faster operation.

*Base offset compression.* In order to store the entire $L_j$ sequence, we may need a huge (say 8-byte) integer number. On the other hand, the sequence is strictly increasing:

$$L_0 < L_1 < ... < L_{N-1} \tag{7}$$

The difference sequence, $\Delta L_j$, contains significantly smaller values. Based on this observation, we may compress the header further.

Suppose that we need $\iota$ bytes to store one element of the $L_j$ sequence. In addition, there exists a natural number $l$ such that for all $k = 0, 1, 2, ...$ the

$$L_{(k+1)l-1} - L_{kl} \tag{8}$$

values may be stored in $\theta$ bytes and $\theta < \iota$. In this case, we can store two sequences instead of $L_j$:

(1) $L_0, L_l, L_{2l}, L_{3l}, ..., L_{\lfloor \frac{N-1}{l} \rfloor l}$

(2) $L_0 - L_0, L_1 - L_0, ..., L_{l-1} - L_0,$
    $L_l - L_l, L_{l+1} - L_l, ..., L_{2l-1} - L_l,$
    ...,
    $L_{\lfloor \frac{N-1}{l} \rfloor l} - L_{\lfloor \frac{N-1}{l} \rfloor l}, L_{\lfloor \frac{N-1}{l} \rfloor l+1} - L_{\lfloor \frac{N-1}{l} \rfloor l}, ..., L_{N-1} - L_{\lfloor \frac{N-1}{l} \rfloor l}$

where $\lfloor x \rfloor$ means the integer part (floor) of $x$: $\lfloor x \rfloor = max\{y \mid y \leq x$ and $y$ is integer$\}$.

**Definition.** Sequence (1) will be called the base sequence, whereas sequence (2) is going to be the offset sequence:

$$B_k = L_{kl} \tag{9}$$

$$O_j = L_j - B_{\lfloor \frac{j}{l} \rfloor} \tag{10}$$

where $k = 0, ..., \lfloor \frac{N-1}{l} \rfloor$ and $j = 0, ..., N-1$. The compression method based on these two sequences will be called the base-offset compression (BOC). The base and the offset sequences together will be called the BOC header. □

From the definition of the offset sequence, the following formula for the logical position follows immediately:

$$L_j = B_{\lfloor \frac{j}{l} \rfloor} + O_j \tag{11}$$

Now, let us compare the size of the LPC header to the BOC header. One element of the base sequence occupies $\iota$ bytes, whereas one offset sequence element needs $\theta$ bytes. Thus the space requirements of the two techniques is the following:

LPC: $N\iota$

BOC: $(\lfloor \frac{N-1}{l} \rfloor + 1)\iota + N\theta$

The question is when the header with BOC is smaller than with LPC? In order to give a simple sufficient condition, let us estimate the size of the BOC header from above as follows:

$$\left(\left\lfloor \frac{N-1}{l} \right\rfloor + 1\right)\iota + N\theta \leq \left(\frac{N-1}{l} + 1\right)\iota + N\theta <$$

$$< \left(\frac{N}{l} + 1\right)\iota + N\theta = N\left(\frac{\iota}{l} + \frac{\iota}{N} + \theta\right)$$

We obtained a sufficient condition:

$$N\left(\frac{\iota}{l} + \frac{\iota}{N} + \theta\right) < N\iota \tag{12}$$

$$\frac{\iota}{l} + \frac{\iota}{N} + \theta < \iota \tag{13}$$

That is, if (1.13) holds, then the header with BOC will be smaller than with LPC. $\frac{\iota}{N}$ tends to 0, if $N$ tends to $\infty$. Therefore, for sufficiently large $N$ values, it is enough to check the following approximate inequality:

$$\frac{\iota}{l} + \theta < \iota \tag{14}$$

In case of the TPC-D benchmark database, a suitable value of $l$ was 64, with $\iota = 8$ and $\theta = 4$. The header decreased from 45.8 Megabytes to 23.6 Megabytes, to its 51.6%. The left side of inequality (1.14) divided by the right side gives approximately the same result:

$$\frac{\frac{\iota}{l} + \theta}{\iota} \approx 51.6\% \tag{15}$$

which proves the usability of the this approximate inequality.

We could decrease the size of the header in the TPC-D benchmark database to its 1/2 by applying LPC instead of SCHC. On the other hand, the header with BOC is 51.6% of the header with LPC. That is the original header was compressed to its 25.8% (from 91.6 Megabytes to 23.6 Megabytes). The base array was small, it occupied 732.5 Kilobytes of memory only. The offset array was 22.9 Megabytes. Both the base and the offset arrays fit into the physical memory. If not, then the binary search may be implemented in such a way that the virtual memory pagings are minimized:

(1) First, the binary search is done on the base array.

(2) Then, when the adequate $l$-long section is found, the binary search is continued in the offset array.

# 4 Comparison of Physical Database Representations

In addition to the empirical results mentioned in the previous section, it is also possible to prove that, in the worst case, the multidimensional representation of a relation with SCHC occupies less space than the table representation of the same relation. (Here the multidimensional representation consists of the compressed array and the SCHC header, the table representation includes the table plus the B-tree.)

**Definition.** In this section, we are going to use the following notations:
$\delta$ = data ratio in the table ($0 \leq \delta < 1$); it equals the total size of the columns outside the key divided by the size of the row;
$\theta$ = size of the record identifier (RID) in the B-tree ($\theta > 0$);
$S$ = size of one row in the table ($S > 0$);
$\iota$ = size of the one-dimensional index ($\iota > 0$);
$N$ = number of elements in relation $R \subseteq D_1 \times ... \times D_n$ ($N > 0$);
$C_T$ = cost of table-representation ($C_T > 0$);
$C_M$ = cost of multidimensional array representation ($C_M > 0$). □

**Costs.** The cost of table representation will include the size of the table ($NS$) plus the size of the B-tree. In the worst case scenario, $C_T$ will equal the following expression:

$$C_T = NS + 2(1 - \delta)NS + 4N\theta \qquad (16)$$

In the worst case, the B-tree is two times as big as in the best case. In the best case, almost all pages of the B-tree are full. These pages store, among others, the keys and RIDs of rows corresponding to the keys. In addition, the pages have to store the RIDs of their children. The size of all keys is $(1-\delta)NS$. The size of all RIDs (including the RIDs of rows and RIDs of children) is $2N\theta$. That is why the size of the B-tree may be estimated with $(1 - \delta)NS + 2N\theta$ in the best case. In the worst case, its estimation is $2(1 - \delta)NS + 4N\theta$.

The cost of multidimensional array representation will consist of the size of the compressed array ($\delta NS$) plus the size of the header. In the worst case, it will equal:

$$C_M = \delta NS + 2N\iota \qquad (17)$$

The compressed array does not store the keys, that is why there is a $\delta$ coefficient in the first member of the sum. In the worst case, the number of runs equals $N$. Thus the size of the SCHC header is $2N\iota$.

**Remark.** The cost definitions do not contain all the possible cost elements, only the most important ones. For example, the B-tree pages store a flag as well, which shows whether the page is a leaf page or not. In the multidimensional representation, the dimension values have to be stored, too.

**Theorem.** The multidimensional database organization with single count header compression occupies less space than the table-based organization with B-tree.

**Proof.** The worst cases will be compared with each other. Consider the physical representations of relation R. The size of the table representation of R is at most $C_T$, whereas the size of the multidimensional array representation is at most $C_M$. Let us take the quotient of these two positive numbers:

$$\frac{C_T}{C_M} = \frac{NS + 2(1-\delta)NS + 4N\theta}{\delta NS + 2N\iota} = \frac{3 - 2\delta + 4\frac{\theta}{S}}{\delta + 2\frac{\iota}{S}} \qquad (18)$$

We are going to prove that $\frac{C_T}{C_M} > 1$. The following equation holds:

$$\frac{\iota}{S} \leq 1 - \delta \qquad (19)$$

It is true, because $\iota$ is the size of the one-dimensional index and therefore it must not be larger than the size of the key, which is $(1-\delta)S$. We may think of the one-dimensional index as a compression of the key.

Since $\frac{\iota}{S} \leq 1 - \delta$, the denominator can be increased as follows:

$$\frac{3 - 2\delta + 4\frac{\theta}{S}}{\delta + 2\frac{\iota}{S}} \geq \frac{3 - 2\delta + 4\frac{\theta}{S}}{\delta + 2(1-\delta)} = \frac{3 - 2\delta + 4\frac{\theta}{S}}{2 - \delta} \qquad (20)$$

Because $\delta < 1 < 1 + 4\frac{\theta}{S}$ ($\delta < 1$, $\theta$ and $S$ are positive), we can decrease the counter in the following way:

$$\frac{3 - 2\delta + 4\frac{\theta}{S}}{2 - \delta} = \frac{2 - 2\delta + 1 + 4\frac{\theta}{S}}{2 - \delta} > \frac{2 - 2\delta + \delta}{2 - \delta} = 1 \qquad (21)$$

We obtained that $\frac{C_T}{C_M} > 1$, which means that the table representation with B-tree occupies more space than the multidimensional database organization with single count header compression. ∎

In order to visualize the ratio $\frac{C_T}{C_M}$, let us put these values into a table (see Table 1).

$\theta/S = 20\%$ means that the record identifiers of the B-tree are 1/5 of the length of a row in the table. Similarly, $\iota/S$ compares the length of the one-dimensional index to the length of a row. $\delta$ denotes the data ratio, that is the size of the non-key columns divided by the size of a row.

## 5 Experiments

Experiments were made to verify the theoretical results. Two benchmark databases were tested:

Table 1: $\theta/S = 20\%$

|  | $\iota/S$ | | | | | |
| ---: | ---: | ---: | ---: | ---: | ---: | ---: |
| $\delta$ | 0% | 20% | 40% | 60% | 80% | 100% |
| 0% | - | 9.50 | 4.75 | 3.17 | 2.38 | 1.90 |
| 20% | 17.00 | 5.67 | 3.40 | 2.43 | 1.89 | - |
| 40% | 7.50 | 3.75 | 2.50 | 1.88 | - | - |
| 60% | 4.33 | 2.60 | 1.86 | - | - | - |
| 80% | 2.75 | 1.83 | - | - | - | - |
| 100% | 1.80 | - | - | - | - | - |

- the TPC-D benchmark database [8];

- and the APB-1 benchmark database [4].

The specifications of both benchmark databases can be downloaded freely from the Internet. (See the URLs in the **References**.) Moreover, the specifications include programs, which are able to generate the benchmark databases.

The TPC-D benchmark database was prepared by the program called DBGEN. The size of the created database was 1 GB (the scale factor given to DBGEN was equal to 1). Then a relation was derived from the database with three dimensions: Product, Supplier and Customer. One measure attribute (Extended Price) was chosen for testing purposes. A similar relation was used in [2], [3] and [7].

The APB-1v2 File Generator obtained three parameters: It used 10 channels, the density was 1%, and it assumed 10 users. From the created relations, we worked on the one, which was stored in the file called *histsale.apb*. The four dimensions (Customer, Product, Channel and Time) and one measure attribute (Dollar Sales) were kept and used in the testing.

The table-representation of relation R consists of one table and a B-tree index. The multidimensional representation of the same relation will be constituted by the following things:

- The compressed array;

- The base and the offset arrays of the header;

- One array per dimension to store the dimension values.

The space requirements of the two physical representations of the TPC-D database and the APB-1 database are described in Table 2 and Table 3. (The description of hardware and software, which were used during testing, can be found in the **Appendix**.)

It is interesting to compare the size of the multidimensional representation with the size of the table representation compressed with different software products. This comparison can be found in Table 4 for the TPC-D benchmark database, whereas for the APB-1 benchmark database in Table 5.

Table 2: Table representation

| File | Size in Bytes TPC-D | Size in Bytes APB-1 |
|---|---:|---:|
| Table | 120,019,300 | 644,436,000 |
| B-tree index | 159,617,024 | 650,792,960 |
| Total | 279,636,324 | 1,295,228,960 |

Table 3: Multidimensional representation

| File | Size in Bytes TPC-D | Size in Bytes APB-1 |
|---|---:|---:|
| Compressed array | 48,007,720 | 99,144,000 |
| Base array | 750,128 | 1,549,128 |
| Offset array | 24,003,860 | 24,786,000 |
| Dimension 1 | 800,000 | 8,320 |
| Dimension 2 | 40,000 | 84,500 |
| Dimension 3 | 399,984 | 117 |
| Dimension 4 |  | 119 |
| Total | 74,001,692 | 125,572,184 |

The size of the uncompressed table-representation was minimal, because the B-tree was completely saturated. (The compete saturation was achieved in the following way. First the table was ordered by the key. Then the records with odd record number were added to the B-tree in increasing order. This resulted in a completely unsaturated B-tree. Finally, the records with even record number were added to the B-tree in increasing order.) Despite this fact, the compression programs could decrease the size of the table representation to 1/3 of its original size in the TPC-D database and to its 1/10 in the APB-1 database.

The best one could decrease the size of the TPC-D database table representation to its 29%. The multidimensional representation is only 26% of the table representation. The multidimensional representation may be considered as another type of compression, which is better than all the other examined programs because

- it results in a smaller database size
- and it allows faster access to the cells than the uncompressed table representation.

In case of the APB-1 benchmark database, we obtain a slightly different result: The size of the multidimensional representation is not smaller than the compressions of the table representation, but it is comparable with them.

Table 4: TPC-D benchmark database

| Compression | Size in Bytes | Percentage |
|---|---|---|
| Uncompressed table representation | 279,636,324 | 100% |
| ARJ | 92,429,088 | 33% |
| gzip | 90,521,974 | 32% |
| WinZip | 90,262,164 | 32% |
| PKZIP | 90,155,633 | 32% |
| jar | 90,151,623 | 32% |
| bzip2 | 86,615,993 | 31% |
| WinRAR | 81,886,285 | 29% |
| Multidimensional representation | 74,001,692 | 26% |

Table 5: APB-1 benchmark database

| Compression | Size in Bytes | Percentage |
|---|---|---|
| Uncompressed table representation | 1,295,228,960 | 100% |
| jar | 124,462,168 | 10% |
| gzip | 124,279,283 | 10% |
| WinZip | 118,425,945 | 9% |
| PKZIP | 117,571,688 | 9% |
| ARJ | 115,085,660 | 9% |
| bzip2 | 99,575,906 | 8% |
| WinRAR | 98,489,368 | 8% |
| Multidimensional representation | 125,572,184 | 10% |

The speed of the retrieval operation was also tested. Random samples were taken from the relation. The elements of the sample were sought in the table representation one by one through the B-tree. Then the same sample elements were sought in the multidimensional representation, as well, with the help of the BOC header. The sample size was 100, 500, 1,000, 5,000, 10,000, 50,000 and 100,000 in the experiments. Table 6 and Table 7 show the results of these tests.

In column 2 and 3 of Table 6 and Table 7, the length of the retrieval operation is shown in seconds. The last column gives the quotient of the second and the third columns. It says that the multidimensional representation with BOC results in 1.4 - 7.8 times faster operation than the table representation depending on the sample size. The quotient as a function of sample size is drawn in Figure 1.

ON THE SCALABILITY OF MULTIDIMENSIONAL DATABASES    15

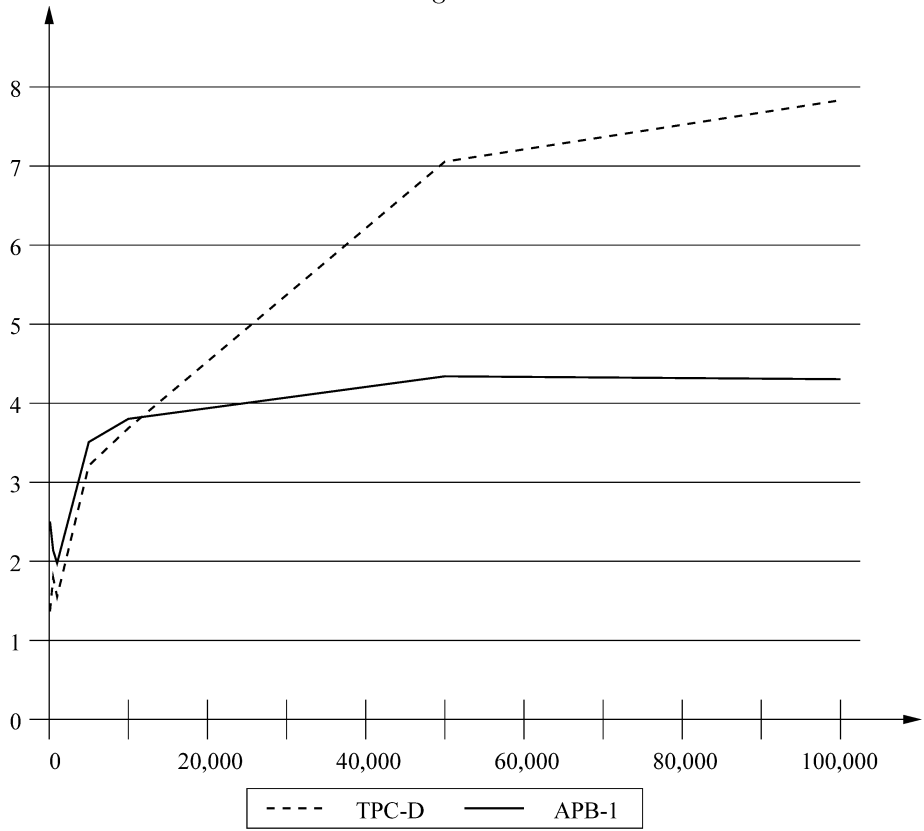

Figure 1:

Table 6: Speed of retrieval in the TPC-D benchmark database

| Sample Size | Table Representation | Multidimensional Representation | Quotient |
|---:|---:|---:|---:|
| 100 | 2.90 | 2.12 | 1.37 |
| 500 | 8.91 | 4.97 | 1.79 |
| 1000 | 10.93 | 7.04 | 1.55 |
| 5000 | 164.86 | 51.35 | 3.21 |
| 10000 | 321.11 | 87.17 | 3.68 |
| 50000 | 1568.72 | 222.36 | 7.05 |
| 100000 | 3148.17 | 402.02 | 7.83 |

Table 7: Speed of retrieval in the APB-1 benchmark database

| Sample Size | Table Representation | Multidimensional Representation | Quotient |
|---:|---:|---:|---:|
| 100 | 6.36 | 2.54 | 2.50 |
| 500 | 15.63 | 7.31 | 2.14 |
| 1000 | 28.73 | 14.53 | 1.98 |
| 5000 | 230.13 | 65.59 | 3.51 |
| 10000 | 444.74 | 116.97 | 3.80 |
| 50000 | 2214.60 | 510.21 | 4.34 |
| 100000 | 4450.14 | 1033.93 | 4.30 |

# 6 Conclusion

There are several different definitions of scalability. If an on-line analytical processing (OLAP) tool is strong according to one definition, it may be still weak according to some others.

If we take the database size as the primary criterion for scalability, then the compressed multidimensional physical database representation may be more scalable than the table-based one, because the former one results in smaller database size.

In many OLAP application, the stored data is used in a read-only or read-mostly manner. The database of lot of application is refreshed only periodically (daily, weekly, monthly, etc.). In these applications, it is acceptable, that the data is loaded and compressed using batch processing outside working hours. Moreover, it does not make any difficulty in SCHC, LPC and BOC, if we want to update an already existing (that is nonempty) cell.

On the other hand, if we want to fill in an empty cell or empty a nonempty one, then we have to insert a new cell in the compressed array or delete an existing one from it, which is a much more expensive operation. The relational databases were designed to cope with large number of update, insert and delete transactions. That is why it may be more beneficial to use the table representation, if we have to insert or delete frequently.

Speed is extremely important in the field of OLAP. Memory operations are faster than hard disk operations often with more orders of magnitude. If we can compress the OLAP database so that the entire database fits into the (physical) memory, then we may be able to speed up the OLAP transactions significantly. This is another reason, why it is advantageous to find better and better compression techniques.

## Acknowledgments

I would like to thank Prof. Dr. János Csirik for his invaluable comments on earlier versions of this paper, Mr. Nigel Pendse for sending me his paper entitled Scalability.

## Appendix

The table below shows the hardware and software, which were used for testing.

| | |
|---|---|
| Computer | Toshiba Satellite 300CDS |
| Processor | Intel Pentium MMX |
| Processor speed | 166 MHz |
| Memory size | 80 MB |
| Hard disk manufacturer | IBM |
| Hard disk size | 11 GB |
| File system | ext2 |
| Page size of B-tree | 4 KB |
| Operating system | Red Hat Linux release 6.2 (Zoot) |
| Kernel version | 2.2.14-5.0 |
| Compiler | gcc version egcs-2.91.66 19990314/Linux |
| Programming language | C |